%% file: 0work.tex
\documentclass[twoside]{article}

\usepackage[letterpaper,hmargin={0.7in,0.6in},vmargin={0.7in,0.9in}]{geometry}	
\usepackage[lite]{amsrefs}															
\usepackage{amsmath}
\usepackage{amssymb,latexsym}													
\usepackage[latin1]{inputenc}													
\usepackage{bm} 																
\usepackage{multicol}															
\renewcommand{\columnseprule}{.4pt}												
\usepackage{feynmf}																
\usepackage{mathrsfs}															
\usepackage{graphicx}
\usepackage{color}

\newcommand{\ud}{\mathrm{d}}
\newcommand{\uD}{\mathscr{D}}
\newcommand{\x}{\mathbf{x}}

\newcommand{\p}{\mathbf{p}}

\newcommand{\w}{\omega}

\begin{document}
\title{Order of the chiral phase transtion in the $\sigma$ models}
\author{F. A. Correa$^{1,}$\thanks{facorread@unal.edu.co}
\\J. Morales$^{1,2,}$\thanks{jmoralesa@unal.edu.co}
	\\\emph{\normalsize $^1$Departamento de Física, Universidad Nacional, Bogotá, Colombia.}
	\\\emph{\normalsize $^2$Centro Internacional de Física, Bogotá, Colombia.}
	}
\date{\today}
\maketitle
\begin{abstract}
The \emph{chiral phase transition} at a certain \emph{critical temperature} is a restoration mechanism of the chiral symmetry, broken by the nonzero mass of quarks and mesons. The transition can be studied through several models, among which are the $\sigma$ models. An analysis is made on the linear and nonlinear $\sigma$ models with different approximations, and we show that the transition is of second order in both cases.
\end{abstract}

\begin{fmffile}{fabio} 
\begin{multicols}{2}
\section{Introduction}
A \emph{chiral transformation} is a rotation in the \emph{isospin} space of quarks. When a quark system is invariant under such a transformation, we
say that the system is chirally symmetric. This is true in the case of massless quarks, but we know that actually they have mass. So with regard to the light quarks
only (\emph{up} and \emph{down}), we can say that the chiral symmetry is an approximated symmetry of the strong interactions \cite{koch}. Nonzero mass of the
corresponding mesons also breaks the symmetry.

There are two kinds of chiral transformations. The partially conserved Noether current is a vector in one case, and an axial vector in the other case. When working
with the up and down quarks, the transformations can be studied through the group SU(2)$\times$SU(2), which is isomorphic to O(4). This isomorphism is the reason for
the Heisenberg magnet model being the unique description of the low energy dynamics of the QCD.

The $\sigma$ models arise in this context. The $\sigma$ model has the usual kinetic term and the potential

\begin{equation}
	\frac{\lambda}{4}\left(\mathbf{\Phi}^2 -f_\pi^2 \right)^2,
	\end{equation}
with a field $\mathbf{\Phi}$ having $N$ components, where $\lambda$ is a positive coupling constant, and $f_\pi$ is the pion decay constant; this model is
renormalizable. In the limit that $\lambda\rightarrow\infty$ the potential goes over to a $\delta$-function constraint on the length of the field vector. This is the
so-called nonlinear $\sigma$ model. Both models describe a quark system in which the chiral symmetry is broken, in consistency with the nonzero mass of the light
quarks and/or the light mesons.

If quarks are massless, QCD is expected to undergo a \emph{chiral phase transition}, which may have implications for high energy nucleus-nucleus collisions. A pion gas
with a $\sigma$ meson undergoes the chiral phase transition. In this work, we use both the $\sigma$ models to describe it and derive its pressure. The linear $\sigma$
model is studied through a mean field approximation, while the nonlinear $\sigma$ model is studied using chiral perturbation theory. The first and second term of the
perturbation series are considered. In both studies, the transition is of second-order and the critical temperature is the same, as we will show later.

The developments towards the calculation of the critical temperature follow closely those of Bochkarev and Kapusta\cite{bochkarev}.

\section{Linear sigma model}
The linear $\sigma$ model Lagrangian is
\begin{equation}
	\mathscr{L} = \frac{1}{2}\left( \partial_\mu\mathbf{\Phi}^2 \right)^2 -\frac{\lambda}{4}\left(\mathbf{\Phi}^2 - f_\pi^2\right)^2,
	\end{equation}
where $\lambda$ is a positive coupling constant. The bosonic field $\mathbf{\Phi}$ has $N$ components. Rather arbitrarily, we define that they are $N-1$ $\pi$
fields and a $\sigma$ field as the $N$th component. The space spanned by these components has the usual Euclidean metric. Since the $\sigma$ has the quantum
number of the vacuum, and since the symmetry is broken at low temperatures, we immediately allow for a $\sigma$ condensate $v$ whose value depends on the
temperature $1/\beta$ and is yet to be determined. 

We write
\begin{gather}
	\Phi_i(\mathbf{x},t) = \pi_i(\mathbf{x},t), \qquad i = 1,\ldots , N-1; \\
	\Phi_N(\mathbf{x},t) = v + \sigma(\mathbf{x},t).
	\end{gather}
\end{multicols}

In terms of these fields the Lagrangian becomes 
\begin{align}
	\mathscr{L} &= \tfrac{1}{2}\left(\partial_\mu\pi\right)^2 +\tfrac{1}{2}\left(\partial_\mu\sigma\right)^2 -\tfrac{\lambda}{4}\left(v^2 -f_\pi^2\right)^2 -\tfrac{\lambda}{2}\left(v^2
		-f_\pi^2\right)\left(2v\sigma +\sigma^2 +\pi^2\right) -\tfrac{\lambda}{4}\left(2v\sigma +\sigma^2 +\pi^2 \right)^2 
	\intertext{We define the effective masses \parbox{3in}{$\begin{cases}\bar{m}_\pi^2 =\lambda \left(v^2-f_\pi^2\right)\\ \bar{m}_\sigma^2 =\lambda \left(3v^2-f_\pi^2\right). \end{cases}$}}
	\intertext{Reorganizing terms,}
	\mathscr{L} &= -\tfrac{\lambda}{4}\left(f_\pi^2 -v^2\right)^2 +\tfrac{1}{2}\left[\left(\partial_\mu\pi\right)^2 -\bar{m}_\pi^2\pi^2 +\left(\partial_\mu\sigma\right)^2
		-\bar{m}_\sigma^2\sigma^2\right] -\lambda v\left(v^2 -f_\pi^2\right)\sigma -\lambda v\sigma\left(\sigma^2 +\pi^2\right) -\tfrac{\lambda}{4}\left(\sigma^2 +\pi^2\right)^2
	\end{align}

\begin{multicols}{2}
At zero temperature, the potential is minimized when $v=f_\pi$. In this particular case,
\begin{align}
	\bar{m}_\pi^2 		&= \lambda \left(v^2-f_\pi^2\right) = 0, \label{LSM-pion_eff_mass}\\
	\bar{m}_\sigma^2	&= \lambda \left(3v^2-f_\pi^2\right) = 2\lambda f_\pi^2,
	\end{align}
so the Goldstone theorem is satisfied.

The action at finite temperature is obtained by rotating to imaginary time, $\tau=it$, and integrating $\tau$ from 0 to $\beta=1/T$. We keep the Minkowsi metric: $\partial_\mu =\partial /\partial
x^\mu$ with $\partial_0 =\partial/\partial t =i\partial/\partial\tau$. Then the action is
\begin{multline}
	S =-\tfrac{\lambda}{4}\left(f_\pi^2 -v^2\right)^2\beta V +\int\ud^4x \biggl( \tfrac{1}{2}\Bigl[\left(\partial_\mu\pi\right)^2 -\bar{m}_\pi^2\pi^2 \\
	+\left(\partial_\mu\sigma\right)^2 -\bar{m}_\sigma^2\sigma^2\Bigr] -\lambda v\left(v^2 -f_\pi^2\right)\sigma \\
	-\lambda v\sigma\left(\sigma^2 +\pi^2\right) -\tfrac{\lambda}{4}\left(\sigma^2 +\pi^2\right)^2 \biggr), \label{MSL S}
	\end{multline}
where the abbreviation
\begin{equation}
	\int\ud^4x \equiv \int_0^\beta\ud\tau \int_V\ud^3x
	\end{equation}
is used.

At any temperature $v$ is choosen such that $\langle\sigma\rangle=0$. This eliminates any one-particle reducible diagrams in perturbation theory, leaving only one-particle irreducible (1PI)
diagrams. We will allow for $v$ to be temperature dependent. The simplest approximation at finite temperature is the mean field approximation; in this, we will neglect interactions among the
particles or collective excitations. 

The pressure of a free relativistic boson gas can be written in several ways:
\begin{align}
	P_0(T,m) &=-\frac{1}{\beta}\int\frac{\ud^3k}{(2\pi)^3} \ln\left(1-e^{-\beta\w}\right) \\
		&=\int\frac{\ud^3k}{(2\pi)^3}\frac{k^2}{3\w}\frac{1}{e^{\beta\w}-1}.
	\end{align}
%
%
The pressure of the pion gas is given by
\begin{multline}
	P = \frac{\ud}{\ud V}\frac{1}{\beta}\ln Z(\beta) =-\tfrac{\lambda}{4}\left(f_\pi^2 -v^2\right)^2 +VP_0(T,m_\sigma) \\
								+V(N-1)P_0(T,m_\pi)
	\end{multline}

The mean field approximation thus calculated has been a very simple, very good approximation, allowing to analyze lots of phenomena in finite temperature field theory.
It was used in all the pioneering papers.

\subsection{second-order phase transition}
We expect that the temperature rise tends to disorder the condensate until it dissapears. In the linear $\sigma$ model we analyze this phenomena by expanding the free
boson gas pressure about zero mass:
\begin{equation}
	P_0(T,m) =\frac{\pi^2}{90}T^4 -\frac{m^2T^2}{24} +\frac{m^3T}{12\pi} +\ldots \label{P0_expansion}
	\end{equation}
\end{multicols}

The efective masses are proportional to the square root of $\lambda$, and terms with $m^3$ or $\lambda^{3/2}$ are not part of the mean field approximation, so we
neglect the last term of the expansion \eqref{P0_expansion}. The pion gas pressure becomes
\begin{align}
	P(T,v) 	&\approx -\frac{\lambda}{4}\left(f_\pi^2-v^2\right)^2 +\left(\frac{\pi^2}{90}T^4-\frac{T^2}{24}m_\sigma^2\right) 
				+(N-1)\left(\frac{\pi^2}{90}T^4-\frac{T^2}{24}m_\pi^2\right) \\
			&\approx N\frac{\pi^2}{90}T^4 -\frac{\lambda}{2}v^2\left(f_\pi^2 -\frac{N+2}{12}T^2\right) 
				-\frac{\lambda}{4}\left(v^4 +f_\pi^4 -\frac{N}{6}T^2f_\pi^2\right)
	\end{align}

\begin{multicols}{2}
Now we maximize the pressure with respect to $v$:
\begin{gather}
	\frac{\ud P}{\ud v} =\lambda v\left(f_\pi^2 -\frac{N+2}{12}T^2\right) -\lambda v^3 =0\\
	\therefore v^2 =\begin{cases}f_\pi^2 -\frac{N+2}{12}T^2 \\ 0 \end{cases} \label{dP/dv}
	\end{gather}

The pressure must be a maximum because we expect the pion gas to be in a stable or metastable state. We verify the value of the second derivative for both solutions.
\begin{align}
	\frac{\ud^2P}{\ud v^2} 	&=\lambda\left(f_\pi^2 -\frac{N+2}{12}T^2\right) -2\lambda v^2 \\
							&=\begin{cases}
								-\lambda v^2, \\
								\lambda\left(f_\pi^2 -\frac{N+2}{12}T^2\right),	\label{LSM condensado nonulo}
								\end{cases}
	\end{align}
with respect to the cases of Eq. \eqref{dP/dv}; the first case represents a sure maximum. 

The condensate goes to zero at a critical temperature given by
\begin{equation}
	T_c^2 =\frac{12}{N+2}f_\pi^2.
	\end{equation}

Above this temperature the condensate is zero, because thermal fluctuations are very large.

We evaluate Eq. \eqref{LSM condensado nonulo} again:
\begin{align}
	\frac{\ud^2P}{\ud v^2} 	&=\begin{cases}
								-\lambda v^2, 										&v^2=\frac{N+2}{12}\left(T_c^2 -T^2\right)\\
								\lambda\frac{N+2}{12}\left(T_c^2 -T^2\right),		&v^2=0.
								\end{cases}
	\end{align}
In the second case the pressure is a minimum at $T < T_c$, so we drop it here.
\end{multicols}

Now we will show that this is a second-order phase transition, calculating the pressure under the two temperature regimes.
\begin{align}
	P_<(T) 	&=N\frac{\pi^2}{90}T^4 +\frac{\lambda}{2}v^2\frac{N+2}{12}\left(T_c^2 -T^2\right) -\frac{\lambda}{4}v^4 +\frac{\lambda}{4}f_\pi^2\left(\frac{N}{6}T^2 -f_\pi^2\right) \\
			&=N\frac{\pi^2}{90}T^4 +\frac{\lambda}{2}\left(\frac{N+2}{12}\right)^2\left(T_c^2 -T^2\right)^2 
				-\frac{\lambda}{4}\left(\frac{N+2}{12}\right)^2\left(T_c^2-T^2\right)^2 +\frac{\lambda}{4}f_\pi^2\left(\frac{N}{6}T^2 -\frac{N+2}{12}T_c^2\right)\\
			&=N\frac{\pi^2}{90}T^4 +\frac{\lambda}{2}\left(\frac{N+2}{12}\right)^2\left(T_c^2 -T^2\right)^2 
				+\frac{\lambda}{4}f_\pi^2\left(\frac{N}{6}T^2 -\frac{N}{12}T_c^2\right) -\frac{\lambda}{24}f_\pi^2 T_c^2\\
			&=N\frac{\pi^2}{90}T^4 +\frac{\lambda}{2}\left(\frac{N+2}{12}\right)^2\left(T_c^2 -T^2\right)^2 
				+\frac{\lambda N}{48}f_\pi^2\left(2T^2 -T_c^2\right) -\frac{\lambda}{24}f_\pi^2 T_c^2\\
	P_>(T)	&=N\frac{\pi^2}{90}T^4 +\frac{\lambda N}{48}f_\pi^2\left(2T^2 -T_c^2\right) -\frac{\lambda}{24}f_\pi^2 T_c^2
	\end{align}

Evaluating the pressure in the critical temperature,
\begin{equation}
	P_<(T) =N\frac{\pi^2}{90}T_c^4 +\frac{\lambda N}{48}f_\pi^2T_c^2 -\frac{\lambda}{24}f_\pi^2T_c^2 =N\frac{\pi^2}{90}T_c^4 -\frac{\lambda N}{48}f_\pi^2T_c^2 =P_>(T)
	\end{equation}

\begin{figure}[h!] 
{\center \input{1order.pstex_t}

}	
\caption{\label{Order 2 pt}Above the critical temperature, there is only one pressure maximum at $\nu=0$. Below $T_C$, there is a real maximum at $\nu\neq 0$}
\end{figure}
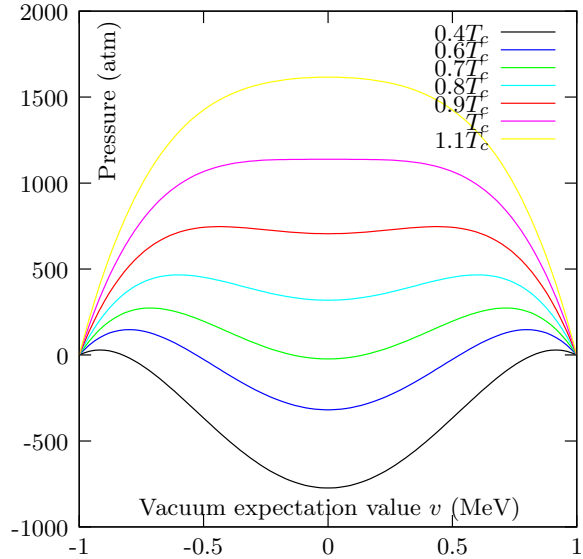

\begin{multicols}{2}
Evaluating the first derivative,
\begin{align}
	P'_<(T) &=N\frac{2\pi^2}{45}T^3 +\lambda\left(\tfrac{N+2}{12}\right)^2\left(T^2-T_c^2\right)T +\frac{\lambda N}{12}f_\pi^2T \notag\\
	P'_>(T) &=N\frac{2\pi^2}{45}T^3 +\frac{\lambda N}{12}f_\pi^2T \\
	P'_<(T_c) &= N\frac{2\pi^2}{45}T_c^3 +\frac{\lambda N}{12}f_\pi^2T_c =P'_>(T_c).
	\end{align}

Evaluating the second derivative,
\begin{align}
	P''_<(T) &=N\frac{2\pi^2}{15}T^2 +\lambda\left(\tfrac{N+2}{12}\right)^2\left(3T^2-T_c^2\right) +\frac{\lambda N}{12}f_\pi^2 \notag \\
	P''_>(T) &=N\frac{2\pi^2}{15}T^2 +\frac{\lambda N}{12}f_\pi^2 \\
	P''_<(T_c) &=N\frac{2\pi^2}{15}T_c^2 +\lambda\left(\tfrac{N+2}{12}\right)^2 2T^2 +\frac{\lambda N}{12}f_\pi^2 \notag \\
		&= P''_>(T_c) +\lambda\left(\tfrac{N+2}{12}\right)^2 2T^2;
	\end{align}
we find that it is discontinuous.

\section{Nonlinear sigma model}
This model is defined by the Lagrangian
\begin{equation}
	\mathscr{L} =\tfrac{1}{2}\left(\partial_\mu\mathbf{\Phi}\right)^2,
	\end{equation}
together with the constraint
\begin{equation}
	f_\pi^2 =\Phi^2(x,t)
	\end{equation}
The length of the chiral field is fixed and cannot be changed by thermal fluctuations, so we say that chiral symmetry is built into this model and, therefore, there can be no chiral-symmetry restoring phase
transition. Taking the limit $\lambda\rightarrow\infty$ constraints the length of the chiral field to be $f_\pi$ just as in the nonlinear model. The critical temperature, however, is independent of $\lambda$ at least
in the mean field approximation. So it would seem that the phase transition survives. If it is true, then we ought to be able to derive it entirely within the context of the nonlinear $\sigma$ model. That is what we
shall do, although it involves a lot more effort than the treatment of the linear model. Since the only parameter in the model is $f_\pi$, and we are interested in temperatures comparable to it, we cannot do an
expansion in powers of $T/f_\pi$. The only parameter is $N$, the number of field components. This suggests an expansion in powers of $1/N$.

We begin by representing the field-constraining $\delta$ function by an integral.
\begin{equation}
	Z =\int\uD\Phi\uD b'\exp\int\ud^4x\,\left(\mathscr{L} +ib'\left(\Phi^2-f_\pi^2\right)\right).\label{NSM Z}
	\end{equation}
The zero frequency, zero momentum condensate $v$ is allowed. Following Polyakov, we also separate the zero frequency, zero momentum mode of the auxiliary field $b'$

Integrating over all the other modes will give us an effective action involving the constant part of the fields. We will then minimize the free energy with respect to
these constant parts, which is a saddle point approximation. Integrating over fluctuations about the saddle point is a finite volume correction and of no consequence
in the thermodynamic limit. The Fourier expansions are
\begin{align}
	\Phi_i(\x,\tau) &= \pi_i(\x,\tau) =\sqrt{\frac{\beta}{V}}\sum_{\p,n}e^{i(\x\cdot\p +\w_n{\tau})}\tilde\pi_i(\p,n),\\
	\Phi_N(\x,\tau) &= v +\sigma(\x,\tau) \\&
	= v +\sqrt{\frac{\beta}{V}}\sum{\p,n}e^{i(\x\cdot\p+\w_n\tau)}\tilde\sigma(\p,n),\\
	b'(\x,\tau)		&= i\frac{m^2}{2} +b(\x,\tau) \\&
	=i\frac{m^2}{2}+T\frac{\beta}{V}\sum_{\p,n}e^{i(\x\cdot\p+\nu_n\tau)}\tilde b(\p,n),\label{NSM fourier b}
	\end{align}
where we define the abbreviation
\begin{equation}
	\sum_{\p,n}	\equiv \sum{n}\int{\ud^3p}{(2\pi)^3}.
	\end{equation}

The zero frequency and zero momentum modes have been excluded from the summations. The field $\Phi$ must be periodic in imaginary time for the usual reasons but there is
no such requirement on $b$, hence $\w_n=2\pi nT$ and $\nu_n=\pi nT$. Since $b$ has dimensions of inverse length squared we inserted another factor of $T$ so as to make
its Fourier amplitude dimensionless, as it is for the other fields. From \eqref{NSM Z}, the action becomes
\begin{align}
	S	&=\int\ud^4x\,\biggl(\tfrac{1}{2}\left[(\partial_\mu\pi)^2 +(\partial_\mu(\nu+\sigma))^2\right] \notag\\ &\hspace{36pt}
	+i\left(\tfrac{m^2}{2}+b\right) \left(\pi^2	+(v+\sigma)^2-f_\pi^2\right)\biggr)\\
	&= \int\ud^4x\,\biggl(\tfrac{1}{2}\left[(\partial_\mu\pi)^2 -m^2\pi^2 +(\partial_\mu\sigma)^2 -m^2\sigma^2\right] \notag\\ &\hspace{18pt}
	-ib(2v\sigma +\pi^2 +\sigma^2)\biggr) +\tfrac{1}{2}m^2(f_\pi^2-v^2)\beta V.
	\end{align}
Please note that terms linear in the fields have been integrated to zero because
\begin{equation}
	\langle\pi_i\rangle =\langle\sigma\rangle =\langle b\rangle =0.
	\end{equation}

An effective action is derived by expanding $e^S$ in powers of $b$. The term linear in $b$ vanishes n account of $\tilde b(0,0)\propto\langle b\rangle =0$. The term
prooprtional to $b^2$ is not zero and is exponentiated, thus summing a whole series of contributions. The term proportional to $b^3$ is not zero either and may also be
exponentiated, summing an infinite series of higher-order terms left out of the order $b^2$ exponentiation. After the expansion is made, we perform a path integration
over the pion and $\sigma$ fields, on ly on the term quadratic in $b^2$. At last, the terms are scaled: $b\rightarrow b/\sqrt{2N}$.

We have
\begin{align}
S_E =&- \frac{1}{2}\sum_{\p,n}\left(\w_j^2+\p^2+m^2\right)\Bigl[\tilde\pi(\p,n)\cdot\tilde\pi(-\p,-n)\notag \\ &
+\sigma(\p,n)\sigma(-\p,-n)\Bigr] -\frac{1}{2}\sum_{\p,n}\biggl[\Pi(\p,\w_n,T,m)\notag \\ &
+\frac{2}{N}\frac{v^2}{\w_n^2+p^2+m^2}\biggr]\tilde b(\p,2n)\tilde b(-\p,-2n)\notag \\ &
+\tfrac{1}{2}m^2(f_\pi^2-v^2)\beta V +\mathscr{O}(b^3/\sqrt{N}).
\end{align}
	
The effective action is an infitine series in $b$. The coefficients are frequency and momentum dependent, arising from one-loop diagrams. In addition, each successive
term is supressed by $1/\sqrt{N}$ compared to the previous one. This is the large $N$ expansion.

The propagators for the pion and $\sigma$ fields are of the usual form
\begin{equation}
\mathsf{D}_\phi^{-1}(\p,\w_n,m) =\w_n^2 +\p +m^2
\end{equation}

and the propagator for $b$ is
\begin{equation}
\mathsf{D}_b^{-1}(\p,\w_n,m) =\Pi(\p,\w_n,T,m) +\frac{2}{N}\frac{v^2}{\w_n^2+\p^2+m^2},
\end{equation}
with $m$ and $v$ to be determined. There appears the one-loop function
\end{multicols}

\begin{equation}
	\Pi(\p,\w_n,T,m) =T\sum_i\int\frac{\ud^3x}{(2\pi)^3}\left((\w_n-\w_l)^2 +(\p-k)^2 +m^2\right)^{-1} \left(\w_l^2 +k^2 +m^2\right)^{-1})
	\end{equation}

Keeping only the terms up to order $b^2$ in $S_E$ (The rest vanishes when $\rightarrow\infty$) allows us to obtain an explicit expression for the partition function and the pressure. This includes the
next-to-leading order in $N$:

\begin{align}
Z 
=& e^{\frac{1}{2}m^2\left(f_\pi^2 -v^2\right)\beta V}\exp\left(-NVT^2\frac{1}{2}\sum_{\p,n}\ln\left(\beta^2(\w_n^2+p^2+m^2)\right)\right)\notag \\
&\hspace{1in}\times\exp\left(-\frac{1}{2}VT^2\sum_{\p,n}\ln\biggl[\Pi(\p,\w_n,T,m) +\frac{2}{N}\frac{v^2}{\w_n^2+p^2+m^2}\biggr]\tilde b(\p,2n)\tilde b(-\p,-2n)\right)\\
P =&\frac{T}{V}\ln Z =\frac{1}{2}m^2\left(f_\pi^2 -v^2\right) -\frac{N}{2}T\sum_{\p,n}\ln\left(\beta^2(\w_n^2+p^2+m^2)\right) -\frac{T}{2}\sum_{\p,n}\ln\biggl[\Pi(\p,\w_n,T,m) +\frac{2}{N}\frac{v^2}{\w_n^2+p^2+m^2}\biggr]
\end{align}

The second term under the last logarithm should and will be set to zero at this order. It may be needed at higher order in the $1/N$ theory to regulate infrared divergences.

The pressure is extremized with respect to $m^2$:
\begin{align}
\frac{\partial P}{\partial m^2}=&\frac{T}{VZ}\int\uD\Phi\uD b'\,\exp\left(\int\ud^4x\left[\mathscr{L}+ib'\left(\Phi^2-f_\pi^2\right)\right]\right)\frac{1}{2}\int\ud^4x\left[-\pi^2-\sigma^2+f_\pi^2-v^2\right]=0\\
=&\frac{T}{V}\frac{1}{2}\left<-\pi^2-\sigma^2+f_\pi^2-v^2\right>=-\left<\pi^2\right>-\left<\sigma^2\right>+f_\pi^2-v^2=0
\end{align}

\begin{multicols}{2}
This condition is equivalent to the thermal average of the constraint:
\begin{equation}
f_\pi^2 = v^2 +\left<\pi^2\right>+\left<\sigma^2\right> =\left<\mathbf{\Phi}^2\right>.\label{NSM constraint thermal average}
\end{equation}

If an approximation to the exact partition function is made, such as the large $N$ expansion, this constraint should still be satisfied. In fact, it will isolate a preferred value for $m$.

To leading order in $N$ we may neglect the term involving $\Pi$ entirely. The pressure is then
\begin{equation}
P = \frac{1}{2}m^2 (f_\pi^2-v^2) +NP_0(T,m).
\end{equation}
This must be a maximum with respect to $v$:
\begin{equation}
\frac{\partial P}{\partial v} =-m^2 v =0,
\end{equation}
which is equivalent to the condition $\langle\sigma\rangle=0$ by means of \eqref{NSM fourier b}. There are two possibilities:
\begin{enumerate}
\item $m=0$: There are massless particles, or Goldstone bosons, and the value of the condensate $v$ is determined by the thermally averaged constraint. This is the \emph{asymmetric} phase.
\item $v=0$: The thermally averaged constraint \eqref{NSM constraint thermal average} is satisfied by a nonzero $T$-dependent mass. There are no Goldstone bosons. This is the \emph{symmetric}, or \emph{symmetry-restored} phase.
\end{enumerate}
Evidently, there is a chiral symmetry-restoring phase transition.

In the leading order of the $1/N$ approximation, the particles are represented by free fields with a potentially $T$-dependent mass $m$. For any free bosonic field $\phi$,
\begin{equation}
\frac{\partial P_0}{\partial m^2}(T,m) =\left<\mathbf{\Phi}^2\right> =\int\frac{\ud^3p}{(2\pi)^3}\frac{1}{\w}\frac{1}{e^{\beta\w}-1},
\end{equation}
with $\w^2=p^2+m^2$. Thus, extremizing the pressure with respect to $m^2$ is equivalent to satisfying the thermally averaged constraint:
\begin{align}
\frac{\partial}{\partial m^2}&\left(-\frac{1}{2}T\sum_{\p,n}\ln\left(\beta(w_n^2+p^2+m^2)\right)\right) \notag\\
=&-\frac{1}{2}T\sum_{\p,n}\frac{1}{\w_n^2+p^2+m^2} =-\frac{1}{2}T\sum_{\p,n}\frac{1}{\w_n^2+\w^2} \notag \\
=&-\frac{1}{2}T\inf\frac{\ud^3p}{(2\pi)^3}\frac{\beta}{2\w}\left(1+2\frac{1}{e^{\beta\w}-1}\right)
\intertext{We renormalize the expression:}
=&-\frac{1}{2}T\int\frac{\ud^3p}{(2\pi)^3}\frac{\beta}{\w}\frac{1}{e^{\beta\w}-1} =-\frac{1}{2}\left<\mathbf{\Phi}^2\right>
\end{align}

Then we have
\begin{align}
\frac{\partial P_0}{\partial m^2}(T,m) =& -\frac{1}{2}\left<\mathbf{\Phi}^2\right> \\
\therefore\frac{\partial P}{\partial m^2} =&\frac{1}{2}\left(f_\pi^2-v^2\right) -\frac{N}{2}\left<\mathbf{\Phi}^2\right> =0\\
\therefore f_\pi^2 =& v^2 +\left<\pi^2\right>+\left<\sigma^2\right> =\left<\mathbf{\Phi}^2\right>(!)
\end{align}

We consider now the two different phases:
\begin{itemize}
\item The mass is zero in the \emph{asymmetric} phase. The constraint is satisfied by a temperature-dependent condensate:
\begin{equation}
	v^2(T) =f_\pi^2 -\frac{NT^2}{12}
	\end{equation}
	This condensate goes to zero at a critical temperature of
\begin{equation}
	T_C^2 =\frac{12}{N}f_\pi^2.
	\end{equation}
	At exactly $T_C$, the thermally averaged constraint is satisfied by the fluctuations of $N$ massless degrees of freedom without the help of a condensate.
\item In the symmetric phase the condensate is zero. The constraint is satisfied by thermal fluctuations alone:
\begin{equation}
	f_\pi^2 =N\left<\mathbf{\Phi}^2\right> =N\int\frac{\ud^3p}{(2\pi)^3}\frac{1}{\w}\frac{1}{e^{\beta\w}-1}.
	\end{equation}
\end{itemize}

At fixed T, thermal fluctuations decrease with increasing mass. The constraint is only satisfied by massless excitations at \emph{one single} temperature, namely $T_C$. At temperatures $T>T_C$ the mass must
be greater than zero. Near $T_C$ the mass should be small, and the fluctuations may be expanded about $m=0$ as
\begin{align}
f_\pi^2 \approx& NT^2\left[\frac{1}{12} -\frac{m}{4\pi T} -\frac{m^2}{8\pi^2T^2}\ln\frac{m}{4\pi T} -\frac{m^2}{16\pi^2T^2}\right]\\
T_C^2 =&\frac{12}{N}f_\pi^2 =12T^2\left(\frac{1}{12} -\frac{m}{4\pi T} -{\ldots} \right) =T^2 -\frac{3 mT}{\pi}
\end{align}

As $T$ approaches $T_C$ from above, the mass approaches zero like
\begin{equation}
m = \frac{\pi}{3T}\left(T^2-T_C^2\right).
\end{equation}

This is a second-order phase transition since there is no possibility of metastable supercooled or superheated states. We will evaluate the order of this transition in the cases
\begin{enumerate}
\item[a)] $m=0$,
\item[b)] $v=0$.
\end{enumerate}

\emph{Function evaluation} Given $m\gg 1$ in the neighborhood of $T_C$ we have for any $T$, and in particular, for $T=T_C$,
\begin{align}
P_a(T) =& NP_0(T,0) =N\frac{\pi^2}{90}T^4\\
P_b(T) =&\frac{1}{2}\frac{\pi^2}{9T^2}\left(T^2 -T_C^2\right)^2f_\pi^2 +NP_0\left(T,\frac{\pi}{3T}\left(T^2 -T_C^2\right)\right) \\
	=&\frac{N}{24}\frac{\pi^2}{9}\left(T^2-T_C^2\right)^2\left(\frac{T_C}{T}\right)^2 +N\biggl(\frac{\pi^2}{90}T^4 \notag \\ &\hspace{0.8in}
		-\frac{T^2}{24}\frac{1}{2}\frac{\pi^2}{9T^2}\left(T^2-T_C^2\right)^2\biggr)\\
	=&N\frac{\pi^2}{90}T^4 +\frac{N}{24}\frac{\pi^2}{9}\left(T^2-T_C^2\right)^2\left(\frac{T_C}{T}\right)^2 \notag \\ &-\frac{1}{2}\frac{N}{24}\frac{\pi^2}{9}\left(T^2 -T_C^2\right)^2
\end{align}

\emph{First derivative evaluation}
\begin{equation}
P'_b(T_C) =N\frac{\pi^2}{90}4T_C^3 =P'_a(T_C)
\end{equation}

\emph{Second derivative evaluation}
\begin{align}
P''_b(T_C) =& N\frac{\pi^2}{90}12T_c^2 -4\frac{N}{24}\frac{\pi^2}{9}\left(3T_C^2 -T_C^2\right) \\
	=& P''_a(T_C) -\frac{N}{3}\frac{\pi^2}{9}T_C^2 
	\end{align}

This is a second order phase transition.

The mass must grow faster than the temperature at very high temperatures in order to keep the field fluctuations fixed and equal to $f_\pi^2$. Asymptotically, the particles move nonrelativistically. This
allows us to compute the fluctuations analytically. We get
\begin{align}
f_\pi^2 =N\left(\frac{T}{2\pi}\right)^{3/2}\sqrt{m}e^{-m/T}
\end{align}

This is a trascendental equation for $m$. IT can also be written as
\begin{align}
m =& T\ln\left(\frac{NT}{2\pi f_\pi}\sqrt{\frac{mT}{2\pi f_\pi^2}}\right)
\sim T\ln\frac{T^2}{T_C^2}
\end{align}

It is rather amusing that, in leading order of the large $N$ approximation, the elementary excitations are massless below $T_C$, become massive above $T_C$, and at asymptotically high temperatures move
nonrelativistically.

The result to first order in $1/N$ provides a good insight into the nature of the two-phase structure of the nonlinear $\sigma$ model, but it is not quite satisfactory for two reasons:
\begin{enumerate}
\item It predicts $N$ massless Goldstone Bosons in the broken symmetry phase when we know there ought to be $N-1$.
\item In this model, $T_C^2 =12f_\pi^2/N$, and in the linear $\sigma$ model it is $T_C^2 =12f_\pi^2/(N+2)$. We expect them to be the same in the limit $\lambda\rightarrow\infty$.
\end{enumerate}

Both these problems can be rectified by inclusion of the second order in $1/N$; this is, the inclusion of the $b$ field.

It is natural to expect that the $b$ field will contribute essentially one negative degree of freedom to the $T^4$ term in the pressure so as to give $N-1$ Gioldstone bosons in the low temperature phase.
Therefore, we move one of the $N$ degrees of freedom and put it together with the $b$ contribution.
\begin{multline}
P =\frac{T}{V}\ln Z =\frac{1}{2}m^2\left(f_\pi^2 -v^2\right) \\ 
-\frac{N-1}{2}T\sum_{\p,n}\ln\left(\beta^2(\w_n^2+p^2+m^2)\right) \\ 
-\frac{T}{2}\sum_{\p,n}\ln\biggl[\beta^2(\w_n^2+p^2+m^2)\Pi(\p,\w_n,T,m)\biggr]
\end{multline}

After the procedure described by Bochkarev and Kapusta\cite{bochkarev}, we do an expansion in $m/T$ as before.
\begin{equation}
P =(N-1)\frac{\pi^2}{90}T^4-\tfrac{N+2}{24}m^2T^2 +\tfrac{1}{2}m^2(f_\pi^2-v^2) +\tfrac{N}{12\pi}m^3T
\end{equation}

We maximize in the high temperature phase where $v=0$:
\begin{align}
\frac{\ud P}{\ud m} = -\tfrac{N+2}{12}m&T^2 +mf_\pi^2 +\tfrac{N}{4\pi}m^2T =0\\
f_\pi^2 =&\tfrac{N+2}{12}T^2 -\tfrac{N}{4\pi}mT\\
=& T^2\left[\tfrac{N+2}{12} -\tfrac{N}{4\pi}\tfrac{m}{T}\right].
\end{align}
This gives the same critical temperature as in the mean field treatment of the linear $\sigma$ model:
\begin{equation}
T_C^2=\frac{12}{N+2}f_\pi^2.
\end{equation}
The mass approaches zero from above like
\begin{equation}
m(T)=\frac{\pi(N+2)}{3NT}(T^2-T_C^2)
\end{equation}

We will evaluate the order of this transition in this order of the $\chi$PT, in the cases
\begin{enumerate}
\item[a)] $m=0$,
\item[b)] $v=0$.
\end{enumerate}

\emph{Function evaluation} Given $m\gg 1$ in the neighborhood of $T_C$ we have for any $T$, and in particular, for $T=T_C$,
\begin{align}
P_a(T) =& (N-1)P_0(T,0) =(N-1)\frac{\pi^2}{90}T^4\\
P_b(T) =&(N-1)\frac{\pi^2}{90}T^4 -\frac{\pi^2}{24}\frac{(N+2)^3}{9N^2}\left(T^2-T_C^2\right)^2 \notag \\
	&+\frac{\pi^2(N+2)^2}{18N^2T^2}\left(T^2-T_C^2\right)^2f_\pi^2 \notag \\
	&+\frac{\pi^2}{27}\frac{(N+2)^3}{12 T^2N^2}\left(T^2-T_C^2\right)^3
\end{align}

\emph{First derivative evaluation}
\begin{equation}
P'_b(T_C) =(N-1)\frac{\pi^2}{90}4T_C^3 =P'_a(T_C)
\end{equation}

\emph{Second derivative evaluation}
\begin{align}
P''_b(T_C) =& N\frac{\pi^2}{90}12T_c^2  -\frac{\pi^2}{24}\frac{(N+2)^3}{9N^2}4 \left(3T^2-T_C^2\right) \\
	=& P''_a(T_C) -\frac{\pi^2}{3}\frac{(N+2)^3}{9N^2}T_C^2
	\end{align}

This is a second order phase transition.

\section{Concluding remarks}
Actually, the chiral phase transition has been studied using three approximations: Mean field approximation on the linear $\sigma$ model, first order of $\chi$PT, and
first and second order of $\chi$PT. We have shown in the three approximations that the transition is of second order, However, the critical temperatures are not the
same, as pointed out for the first order of $\chi$PT. This made the inclusion of the second order necessary.

\nocite{itzykson,bailin,scherer,donoghue,landsman,lebellac}

\end{multicols}

\end{fmffile} 

\renewcommand{\columnseprule}{0pt}

\end{document}

%% file: 1order.pstex_t
\begin{picture}(0,0)%
\includegraphics[type=eps,width=3.2in,height=2.9in]{1order.pstex}%
\end{picture}%
\setlength{\unitlength}{3947sp}%
\begingroup\makeatletter\ifx\SetFigFont\undefined%
\gdef\SetFigFont#1#2#3#4#5{%
  \reset@font\fontsize{#1}{#2pt}%
  \fontfamily{#3}\fontseries{#4}\fontshape{#5}%
  \selectfont}%
\fi\endgroup%
\begin{picture}(3615,3459)(1289,-3905)
\put(1659,-3793){\makebox(0,0)[rb]{\smash{{\SetFigFont{9}{10.8}{\familydefault}{\mddefault}{\updefault}{\color[rgb]{0,0,0}-1000}%
}}}}
\put(1659,-3253){\makebox(0,0)[rb]{\smash{{\SetFigFont{9}{10.8}{\familydefault}{\mddefault}{\updefault}{\color[rgb]{0,0,0}-500}%
}}}}
\put(1659,-2712){\makebox(0,0)[rb]{\smash{{\SetFigFont{9}{10.8}{\familydefault}{\mddefault}{\updefault}{\color[rgb]{0,0,0} 0}%
}}}}
\put(1659,-2172){\makebox(0,0)[rb]{\smash{{\SetFigFont{9}{10.8}{\familydefault}{\mddefault}{\updefault}{\color[rgb]{0,0,0} 500}%
}}}}
\put(1659,-1632){\makebox(0,0)[rb]{\smash{{\SetFigFont{9}{10.8}{\familydefault}{\mddefault}{\updefault}{\color[rgb]{0,0,0} 1000}%
}}}}
\put(1659,-1091){\makebox(0,0)[rb]{\smash{{\SetFigFont{9}{10.8}{\familydefault}{\mddefault}{\updefault}{\color[rgb]{0,0,0} 1500}%
}}}}
\put(1659,-551){\makebox(0,0)[rb]{\smash{{\SetFigFont{9}{10.8}{\familydefault}{\mddefault}{\updefault}{\color[rgb]{0,0,0} 2000}%
}}}}
\put(1726,-3905){\makebox(0,0)[b]{\smash{{\SetFigFont{9}{10.8}{\familydefault}{\mddefault}{\updefault}{\color[rgb]{0,0,0}-1}%
}}}}
\put(2507,-3905){\makebox(0,0)[b]{\smash{{\SetFigFont{9}{10.8}{\familydefault}{\mddefault}{\updefault}{\color[rgb]{0,0,0}-0.5}%
}}}}
\put(3289,-3905){\makebox(0,0)[b]{\smash{{\SetFigFont{9}{10.8}{\familydefault}{\mddefault}{\updefault}{\color[rgb]{0,0,0} 0}%
}}}}
\put(4070,-3905){\makebox(0,0)[b]{\smash{{\SetFigFont{9}{10.8}{\familydefault}{\mddefault}{\updefault}{\color[rgb]{0,0,0} 0.5}%
}}}}
\put(4851,-3905){\makebox(0,0)[b]{\smash{{\SetFigFont{9}{10.8}{\familydefault}{\mddefault}{\updefault}{\color[rgb]{0,0,0} 1}%
}}}}
\put(4307,-682){\makebox(0,0)[rb]{\smash{{\SetFigFont{9}{10.8}{\familydefault}{\mddefault}{\updefault}{\color[rgb]{0,0,0}$0.4T_c$}%
}}}}
\put(4307,-794){\makebox(0,0)[rb]{\smash{{\SetFigFont{9}{10.8}{\familydefault}{\mddefault}{\updefault}{\color[rgb]{0,0,0}$0.6T_c$}%
}}}}
\put(4307,-906){\makebox(0,0)[rb]{\smash{{\SetFigFont{9}{10.8}{\familydefault}{\mddefault}{\updefault}{\color[rgb]{0,0,0}$0.7T_c$}%
}}}}
\put(4307,-1018){\makebox(0,0)[rb]{\smash{{\SetFigFont{9}{10.8}{\familydefault}{\mddefault}{\updefault}{\color[rgb]{0,0,0}$0.8T_c$}%
}}}}
\put(4307,-1130){\makebox(0,0)[rb]{\smash{{\SetFigFont{9}{10.8}{\familydefault}{\mddefault}{\updefault}{\color[rgb]{0,0,0}$0.9T_c$}%
}}}}
\put(4307,-1242){\makebox(0,0)[rb]{\smash{{\SetFigFont{9}{10.8}{\familydefault}{\mddefault}{\updefault}{\color[rgb]{0,0,0}$T_c$}%
}}}}
\put(4307,-1354){\makebox(0,0)[rb]{\smash{{\SetFigFont{9}{10.8}{\familydefault}{\mddefault}{\updefault}{\color[rgb]{0,0,0}$1.1T_c$}%
}}}}
\put(2101,-3661){\makebox(0,0)[lb]{\smash{{\SetFigFont{9}{10.8}{\familydefault}{\mddefault}{\updefault}{\color[rgb]{0,0,0}Vacuum expectation value $v$ (MeV)}%
}}}}
\put(1951,-1561){\rotatebox{90.0}{\makebox(0,0)[lb]{\smash{{\SetFigFont{9}{10.8}{\familydefault}{\mddefault}{\updefault}{\color[rgb]{0,0,0}Pressure (atm)}%
}}}}}
\end{picture}%